\documentclass[a4paper,fleqn,usenatbib,twocolumn]{mnras}

\usepackage{newtxtext,newtxmath}

\usepackage[T1]{fontenc}
\usepackage{ae,aecompl}

\usepackage{graphicx}
\usepackage{amsmath}
\usepackage{amssymb}

\title[New measurements on water ice photodesorption]{New measurements on water ice photodesorption and product formation under ultraviolet irradiation}

\author[G. A. Cruz-Diaz et al.]{
Gustavo A. Cruz-Diaz,$^{1,2,3}$\thanks{E-mail: gustavo.a.cruzdiaz@nasa.gov}
Rafael Mart\'\i{}n-Dom\'enech,$^{1}$
Elena Moreno,$^{1,4}$
\newauthor{ Guillermo M. Mu\~{n}oz Caro$^{1}$
and Yu-Jung Chen$^{5}$}
\\
$^{1}$Centro de Astrobiolog\'{\i}a, INTA-CSIC, Carretera de Ajalvir, km 4, Torrej\'on de Ardoz, 28850 Madrid, Spain\\
$^{2}$Bay Area Environmental Research Institute, Petaluma, CA 94952, USA\\
$^{3}$NASA Ames Research Center, Moffett Field, Mountain View, CA 94035, USA\\
$^{4}$ICMM-2016, Sor Juana In\'es de la Cruz, 3, Cantoblanco, 28049 Madrid, Spain\\
$^{5}$Department of Physics, National Central University, Jhongli City, Taoyuan County 32054, Taiwan
}

\date{Accepted XXX. Received YYY; in original form ZZZ}

\pubyear{2017}

\begin{document}
\label{firstpage}
\pagerange{\pageref{firstpage}--\pageref{lastpage}}
\maketitle

\begin{abstract}
The photodesorption of icy grain mantles has been claimed to be responsible for the abundance of gas-phase molecules toward cold regions. Being water a ubiquitous molecule, it is 
crucial to understand its role in photochemistry and its behavior under an ultraviolet field. We report new measurements on the UV-photodesorption of water ice and its H$_2$, OH, 
and O$_2$ photoproducts using a calibrated quadrupole mass spectrometer. Solid water was deposited under ultra-high-vacuum conditions and then UV-irradiated at various temperatures starting from 8 K with a microwave discharged 
hydrogen lamp. Deuterated water was used for confirmation of the results. We found a photodesorption yield of 1.3 $\times$ 10$^{-3}$ molecules per incident photon for water, and 
0.7 $\times$ 10$^{-3}$ molecules per incident photon for deuterated water at the lowest irradiation temperature, 8 K. The photodesorption yield per absorbed photon is given and 
comparison with astrophysical scenarios, where water ice photodesorption could account for the presence of gas-phase water toward cold regions in the absence of a thermal desorption 
process is addressed. 
\end{abstract}


\section{Introduction}

Water is everywhere. This statement has been proven by different space missions like the Infrared Space Observatory (ISO). Gas phase water emission has been found in contrasting 
astrophysical scenarios like shocks, photodissociation regions (PDRs), high and low-mass protostars (e.g. Cernicharo \& Crovisier 2005). In the other hand, water ice was reported 
to exist in several ISO sources such as embedded young stellar objects (YSOs) and in molecular clouds toward the line of sight of field stars. This offered a characterization 
of ice composition in a variety of lines of sights 
(e.g. Gibb et al. 2004). Herschel Space Observatory observed thousands of YSOs, (Pilbratt et al. 2010). Using PACS (Photodetector Array Camera and Spectrometer) several 
groups detected water emission in objects with different stellar associations, sources, and spatial extensions (Meeus  et  al.  2010; Tilling et al. 2012; Thi et al. 2013; 
Howard et al. 2013; Lindberg et al. 2014; Riviere-Marichalar et al. 2015, 2016; and references therein). Using ALMA (Atacama Large Millimeter/submillimeter 
Array), Cieza et al. (2016) detected indirectly the water snow line of the object V883 Ori, concluding snow lines might have highly dynamical behaviors and they should be considered 
when a disk evolution model is under development.

Water ice controls the efficacy of dust particles and planetesimals aggregation by enhancing their stickiness, see Blum \& Wurm (2008) and references therein.
Water photodesorption plays an important role in protoplanetary disks. Making use of computational models, Akinori et al. (2012) reported that photodesorption holds down 
the ice formation sweeping the snow line notably outwards. In addition, the snow line position can be highly determined by the equilibrium between photodesorption and 
ice formation. Not taking into account the photodesorption effect in models translates into a deeper ice absorption feature than astronomical observations. In the meantime, 
models with photodesorption show a shallow absorption feature nevertheless, the later seem to be a better match, see Honda et al. (2016). 

From the laboratory point of view, water ice has been the subject of study for many years. Greenberg et al. (1980) showed one of the first experiments at low temperature (10 K) 
under high-vacuum conditions (10$^{-8}$ mbar). They deposited an ice mantle made of water, carbon monoxide, ammonia, and carbon dioxide. After UV-irradiation, new molecules 
were produced as a result. Years later, Allamandola et al. (1988) started performing water matrix experiments to recreate a more realistic interstellar ice mantle analog. Westley et 
al (1995a,b) reported the first measurements on water photodesorption by ultraviolet irradiation. Mu\~noz Caro \& Schutte (2003) and  \"Oberg et al. (2010) investigated the effects 
of water concentration on ice mantles and its importance on photochemistry. They found that in water matrix studies the destruction of the molecules under ultraviolet irradiation 
increases with water concentration. This has a direct effect on the molecule lifetime under protostellar conditions. 

In this paper, we report laboratory experiments of pure water ice under UV-irradiation and ultra-high-vacuum (UHV) conditions. Deuterated water was used to confirm the 
data. Here we present measurements of the photodesorption yields of water, deuterated water, and their photoproducts. We provide quadrupole mass spectrometry calibrated data during the 
UV-irradiation and our results are compared with previous experimental and theoretical works. The paper is organized as follows: The methods are explained in Sect. 2. 
We present our results in Sect. 3, followed by a discussion in Sect. 4, and the conclusions are addressed in Sect. 5.

\section{Methods}

The ISAC set-up was used to conduct laboratory experiments of solid water photoprocessing, see Mu\~noz Caro et al. (2010). 
It has a base pressure typically in the range P = 3.0-4.0 $\times$ 10$^{-11}$ mbar and a minimal temperature of 8 K, 
making use of a closed-cycle helium cryostat, which allows us to grow ice layers by deposition of gas species on a cold infrared transparent window. 
The solid samples are processed by UV-radiation and they can be monitored with \emph{in situ} FTIR spectroscopy in transmittance, 
while molecules desorbing to the gas phase are detected by a quadrupole mass spectrometer.

For the experimental simulations here described we used: H$_2$O (liquid), triple distilled, and D$_2$O (liquid), Cambridge Isotope Laboratories, Inc (C.I.L.) 99.9\%. 
In our experiments, the deposition of water vapor is directed using a capillary 
outlet toward the substrate window; this outlet is separated about 2 cm from the substrate. Therefore, most of the water is condensed on the substrate, for more details see 
Mu\~noz Caro et al. (2010). The ice mantles were irradiated using a microwave discharged 
hydrogen flow lamp (MDHL) which has a strong Ly-$\alpha$ and low molecular hydrogen emission simulating the radiation experienced by interstellar ice mantles. The UV-flux is 
$\approx 2 \times 10^{14}$ photons cm$^{-2}$ s$^{-1}$ with a mean energy of 8.6 eV (strong Ly-$\alpha$ centered at 10.2 eV and molecular hydrogen emissions around 7.8 eV) at the sample position (Mu\~noz Caro et al. 2010, Chen et al. 2014, Cruz-Diaz et al. 2014a). 
The water (and deuterated water) ice samples were deposited at 8 K and subsequently warmed up to the irradiation temperature of 30, 50, 70, 
and 90 K. After this, the samples were irradiated with cumulative intervals that led to total irradiation times of 1, 3, 10, 30, and 60 min. 
FTIR spectroscopy was performed after deposition of the ice samples, and between each 
irradiation dose, to monitor the evolution of the original ice component. 
In particular, the column densities of solid water and solid deuterium oxide were calculated from their infrared absorption using the formula: 
$N = \frac{1}{\mathcal{A}} \int_{band} \tau_{\nu} d\nu$, where $N$ is the column density in molecules per cm$^{2}$, $\mathcal{A}$ the band strength in cm molecule$^{-1}$, 
$\tau_{\nu}$ the optical depth of the band, and $d\nu$ the wavenumber differential in cm$^{-1}$. The adopted band strengths were 1.7 $\pm$ 0.2 $\times$ 10$^{-16}$ for the 3259 
cm$^{-1}$ band of H$_2$O (Cruz-Diaz et al. 2014a) and 1.0 $\pm$ 0.2 $\times$ 10$^{-16}$ for the 2413 cm$^{-1}$ band of D$_2$O (Cruz-Diaz et al 2014b). 
Table \ref{exp} summarizes the ice thickness for water and deuterated water experiments, which are only slightly higher than the canonical interstellar ice thicknesses 
(Zubko et al. 2004).

 \begin{table}[ht!]
 \centering
 \caption{Water and deuterated water irradiation experiments performed for this study.}
 \begin{tabular}{ccc}
 \hline
 \hline
 H$_2$O&Irrad. temp.& Thickness\\
 Sample No.&[K]&[ML]\\
 \hline
 1&8&203\\
 2&30&210\\
 3&50&205\\
 4&70&201\\
 5&90&211\\
 \hline
 D$_2$O&&\\
 Sample No.&&\\
 \hline
 6&8&245\\
 7&30&251\\
 8&50&248\\
 9&70&249\\
 10&90&252\\
 \hline
 \end{tabular}
  \\
   One monolayer (ML) corresponds to 1 $\times$ 10$^{15}$ molecules cm$^{-2}$.
 \label{exp}
 \end{table}

The number photons absorbed by the ice thickness can be derived using the Beer-Lambert’s law,
\begin{equation}
I_t = I_0  \: exp(-\sigma N) 
\end{equation}
where $I_{t}$ is the transmitted intensity, $I_{0}$ the incident intensity, $N$ is the ice column density in cm$^{-2}$, and $\sigma$ is the cross section in cm$^{2}$. $N$ was taken 
from Table \ref{exp}, $I_{0}$ is the photon flux, and $\sigma$ the average VUV absorption cross-section of solid H$_2$O or D$_2$O: 3.4 $\times$ 10$^{-18}$ cm$^{2}$ and 
2.7 $\times$ 10$^{-18}$ cm$^{2}$, respectively, see Cruz-Diaz et al. (2014a) and (2014b). We found that on average for the water experiments 50 \% of the photons were absorbed while 
on average for deuterated water was 51 \%.

The main photoproducts during water irradiation, according to Westley et al. (1995a,b), Gerakines et al. (1996), and \"Oberg et al (2009), are H$_2$O$_2$, HO$_2$, O$_2$, OH and H$_2$. 
H$_2$ and O$_2$ do not present IR features but we see them using a quadrupole mass spectrometer (QMS). No IR features corresponding to the remaining photoproducts were detected 
during these experimental simulations. This is an effect of the low conversion rate from water to hydrogen peroxide, ∼ 0.1 \%. This would be translated into a production of $\sim$ 
2 ML of H$_2$O$_2$, close to the detection limit. We cannot measure the amount of hydrogen or oxygen trapped in the ice bulk, but since the former is very volatile and the ladder is 
not produced by a primary photodissociation process of H$_2$O, we assume their concentrations are low and the ice sample can be treated as an almost pure water ice. Therefore, we do 
not find relevant to show the IR spectra collected before and after the UV-photoprocessing because we want to focus on the gas data.

As mentioned above, we used quadrupole mass spectrometry (QMS) to detect the gas-phase molecules that desorbed thermally or by an UV-photon from the ice mantle during the 
irradiation period. QMS measurements were carefully calibrated using the method described in Mart\'\i{}n-Dom\'enech et al. (2015), allowing conversion from the integrated ion 
current of a particular mass fragment A(m/z) to photodesorbed column density N(mol cm$^{-2}$). This method uses additional pure CO ice irradiation experiments to derive a 
proportionality constant k$_{CO}$ between the integrated ion current measured for the m/z = 28 mass fragment of photodesorbing CO molecules and the decrease of the CO ice 
column density measured with the FTIR spectrometer during these additional experiments, which is mainly due to photodesorption.

\begin{equation}
k_{CO} = \frac{A(28)}{N(CO)}.
\end{equation}
This proportionality constant $k_{CO}$ can be used to quantify photodesorption of other species like H$_2$O monitored with a particular mass fragment m/z by applying the different 
factors in the equation:

\begin{eqnarray}
N(mol) &=& \frac{A(m/z)}{k_{CO}} \times \frac{\sigma(CO)}{\sigma(mol)} \times \frac{IF(CO^+)}{IF(z)} \times \nonumber  \\
&\times& \frac{FF(28)}{FF(m)} \times \frac{S(28)}{S(m/z)}
\end{eqnarray}
with $\sigma$ the ionization cross section in the mass spectrometer for the first ionization of CO and the species of interest, knowing that the electron energy for ionization 
in the QMS is 70 eV; $IF$ the fraction of ionized molecules with the 
particular charge of the monitored mass fragment; $FF$ the fraction of molecules of the isotopologue of interest leading to the monitored fragment in the mass spectrometer; and $S$ 
the mass sensitivity of the QMS, which is probed regularly using noble gases. For more details on the calibration process, we refer to Mart\'\i{}n-Dom\'enech et al. (2015). 
For calculation purposes we took: $\sigma_{H_2O}$ = 2.097 $\times$ 10$^{-16}$ cm$^2$, $\sigma_{D_2O}$ = 2.049 $\times$ 10$^{-16}$ cm$^2$ (Straub et al. 1998), 
$\sigma_{D_2}$ = 1.02 $\times$ 10$^{-16}$ cm$^2$ (Rapp \& Englander-Golden 1965, tables 77 and 161 in Celiberto et al. 2001, and Yoon et al. 2010), $\sigma_{OH}$ = 1.85 $\times$ 
10$^{-16}$ cm$^2$ (Deutsch et al. 1997), and $\sigma_{O_2}$ = 2.44 $\times$ 10$^{-16}$ cm$^2$ (Rapp \& Englander-Golden 1965).

\begin{figure}
 \includegraphics[width=\columnwidth]{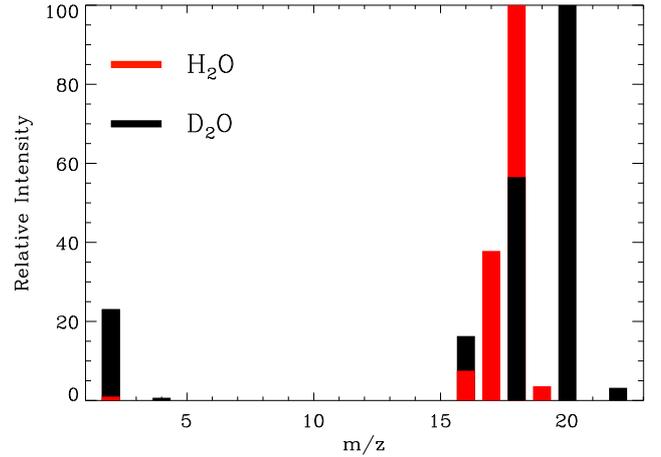}
 \caption{Mass spectrum of water (red bars) and deuterated water (black bars) obtained during deposition of the ice sample using our QMS apparatus at 8 K. The spectra are 
normalized relative to the main mass fragment (m/z = 18 for H$_2$O and m/z = 20 for D$_2$O).}
 \label{masas}
\end{figure}

\section{Experimental results}\label{sec:results}

Figure \ref{masas} shows the mass spectrum of H$_2$O and D$_2$O measured using our QMS apparatus during the deposition of the ice sample. The main mass fragment is 18 for water and 
20 for deuterated water, followed by 17 and 16, and 18 and 16, respectively. H$_2$O and D$_2$O main infrared features are centered around 3290 cm$^{-1}$ and 2470 cm$^{-1}$. 
Although we performed infrared spectroscopy to monitor the deposited ice mantle, we measured the photodesorption yield using our QMS 
because the main band of water and deuterated water are affected by the dissociation of the molecule, in other words, production of OH and OD radicals. 
In particular, the OH absorption feature overlaps with the broad H$_2$O absorption band around 3290 cm$^{-1}$ (O-H stretching mode), and the O$_2$ product is a homonuclear diatomic 
molecule with no dipole which makes it inactive in the infrared.

\subsection{Experiments at 8 K}

\begin{figure}
 \includegraphics[width=\columnwidth]{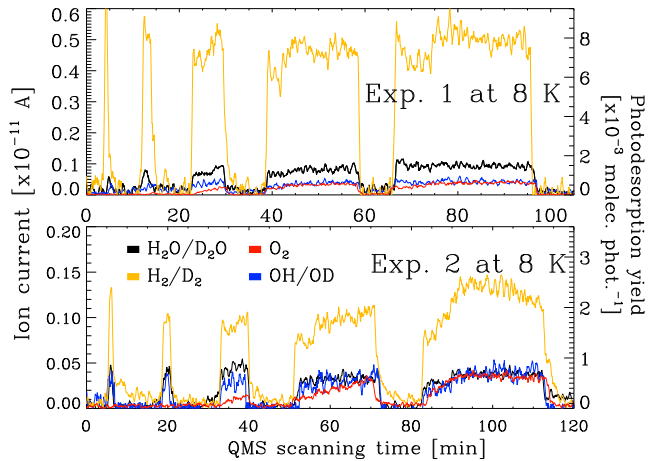}
 \caption{QMS data measuring the photodesorption of different species by UV-irradiation at 8 K. Exp. 1 denotes the irradiation of a pure H$_2$O ice mantle at 8 K while Exp. 2 
 denotes the irradiation of a pure D$_2$O ice mantle at 8 K.}
 \label{8K}
\end{figure}

Experiments were performed at different temperatures as stated previously, but the most relevant temperature to look for a pure photodesorption effect is 8 K. At this 
temperature the only volatile molecules free enough to diffuse through the ice bulk and subsequently desorb thermally are H$_2$ and D$_2$. The influence of heavier molecules 
diffusing throughout the ice bulk could be considered as low since thermal desorption of other detected molecules, O$_2$ and H$_2$O, occurs at higher temperatures, near 30 K and 160 K, 
respectively, (e.g. Collins et al. 2004). Figure \ref{8K} shows the photodesorption observed at 8 K for H$_2$O/D$_2$O, O$_2$, and H$_2$/D$_2$ during the different irradiation intervals.
We were able to measure the photodesorption yield of water and deuterated water deposited at 8 K for different temperatures of irradiation. Table \ref{tablePhoto} shows the average 
measured photodesorption yields at different irradiation temperatures for H$_2$O, D$_2$O, O$_2$, OH, and D$_2$.

\begin{figure}
 \includegraphics[width=\columnwidth]{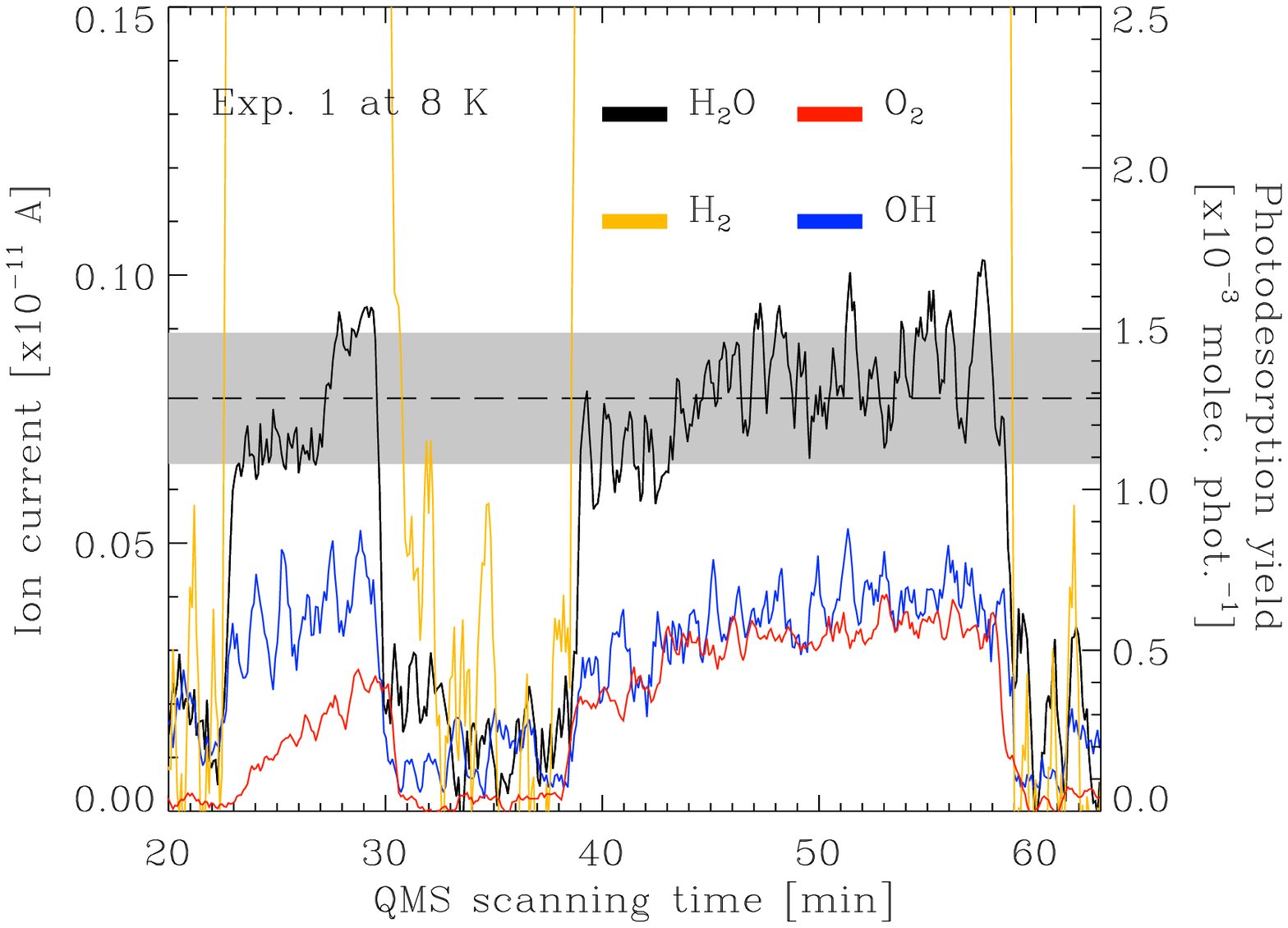}
 \includegraphics[width=\columnwidth]{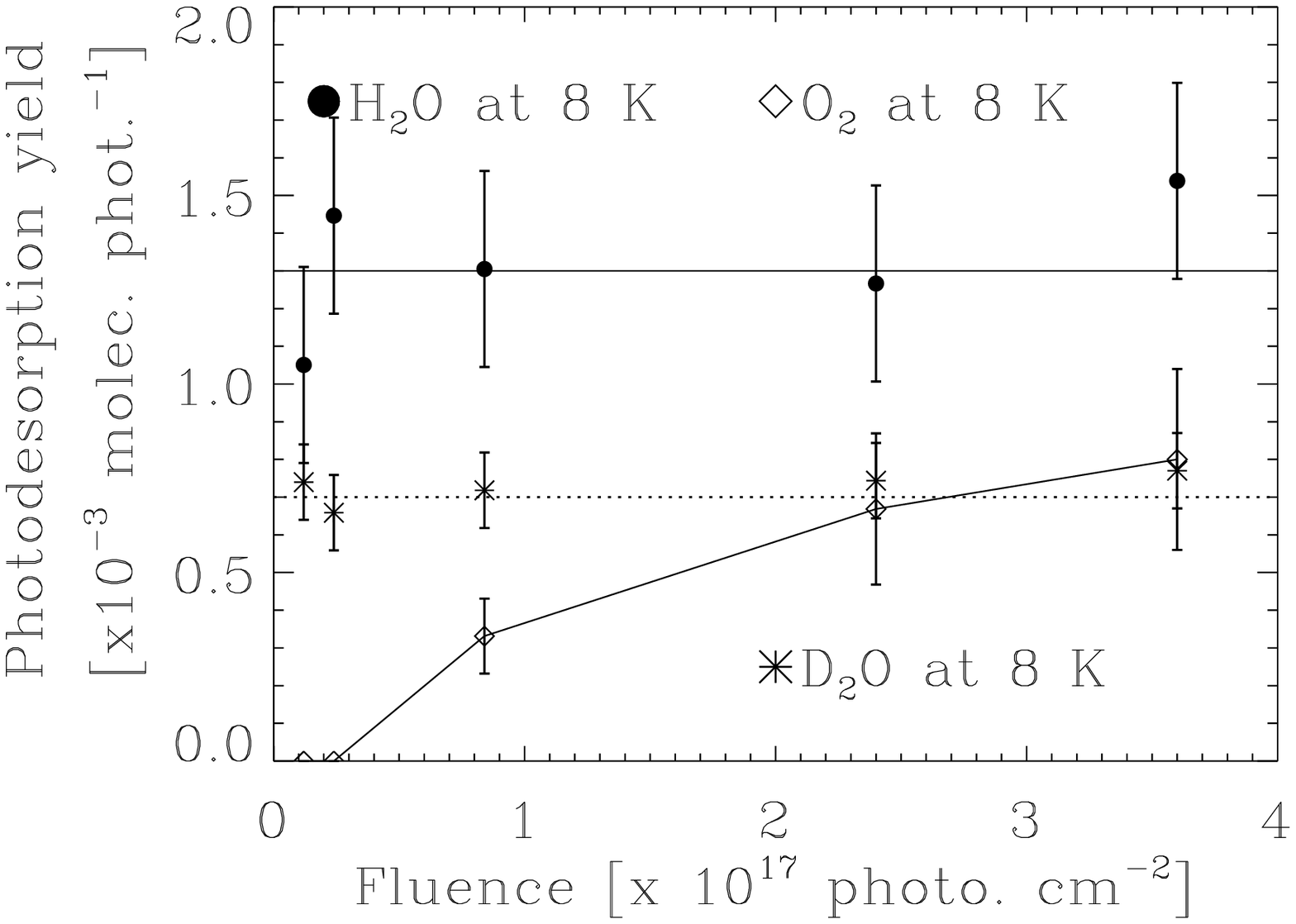}
 \caption{Top: Close-up of Fig. 2 showing the noise in the QMS and the small bumps caused by instabilities in the H$_2$ flux used for the UV-lamp. Exp. 1 denotes the irradiation of a 
 pure H$_2$O ice mantle at 8 K. Bottom: Photodesorption yield as a function of fluence for water (dots), deuterated water (asterisks) and oxygen (diamonds) at 8 K. 
 H$_2$O and D$_2$O present no dependence with the fluence. Solid and dotted  lines represent the average photodesorption yields for water and deuterated water in 
 Table \ref{tablePhoto}.}
 \label{correction}
\end{figure}

Because of fluctuations in the QMS signal (see Fig. \ref{8K} and Fig. \ref{correction} top) we cannot say the H$_2$O/D$_2$O photodesorption is constant. However, we can appreciate 
an average photodeorption for the entire experiment where each irradiation step is within the expected error, see Fig. \ref{correction} bottom. Comparing water and oxygen behaviors 
in this figure, we consider the photodesorption of water and deuterated water at different irradiation periods to be UV-fluence independent. This is not the case for oxygen where 
the photodesorption yield clearly grows with irradiation time. Typically an error of about 12 \% in the photodesorption yields was estimated as the result of UV-flux oscillations 
during the irradiation of the ice and the noise superposed on the QMS signals, gray area in Fig. \ref{correction} top. Although one of the main contaminants inside the chamber is 
water, QMS data was calibrated and water from the background was removed, see Appendix section for a full description of the method.

The OH/OD intensities in the QMS are lower than that of the parent molecule, which leads to a lower signal-to-noise ratio. In the case of OD, its m/z = 18 coincides with 
the molecular mass fragment of the background H$_2$O. The resolution of our QMS does not allow us to distinguish between these two mass fragments, for 
this reason we did not provide the value of its photodesorption yield in Table \ref{tablePhoto}.

Pure CO ice presents 
a constant photodesorption rate for ice thickness above 5 ML (Mu\~noz Caro et al. 2010, Fayolle et al. 2011, Chen et al. 2014). The lower photodesorption rate of H$_2$O did not 
allow the determination of a critical ice thickness. In the case of photoproducts that desorbed either thermally or during irradiation from the ice mantle we detected H$_2$/D$_2$, 
OH/OD, and O$_2$. 

\subsection{\texorpdfstring{H$_2$O/D$_2$O}{LG} photodesorption}

\begin{figure}
 \includegraphics[width=\columnwidth]{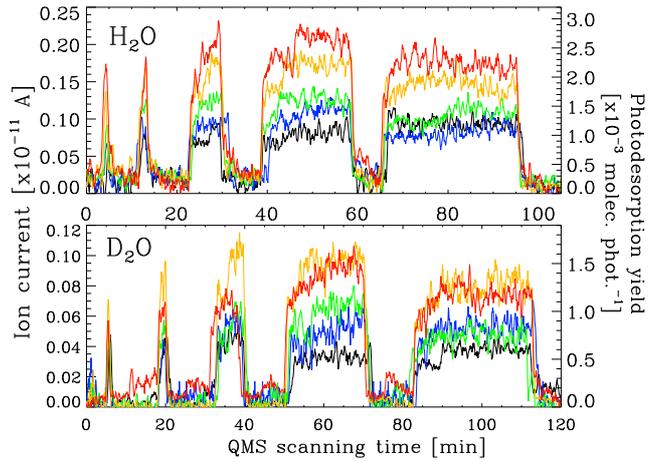}
  \caption{Photodesorption of H$_2$O and D$_2$O deposited at 8 K during UV-irradiation at different temperatures. The different steps follow the irradiation 
  (step) and non-irradiation (flat signal) of the sample by the UV-lamp. This plot follows the color code in Fig. 2. Black = 8 K, blue = 30 K, green = 50 K, yellow = 70 K, and red 
  = 90 K.}
 \label{H2O}
\end{figure}

We performed UV-irradiation of H$_2$O/D$_2$O at different temperatures to study its impact on photodesorption. A previous article reported a similar 
study using methanol and deuterated methanol (see Cruz-Diaz et al. 2016). We found CH$_3$OH and CD$_3$OD photodesorption was too low to detect using our QMS, providing a 
photodesorption upper limit of 3 $\times$ 10$^{-5}$ molecules per incident photon, in line with Bertin et al. (2016). This is not the case for water ice. Figure \ref{H2O} shows the 
QMS signal for water and deuterated water (using their respective main mass fragments) during the ice irradiation at different temperatures. 

\begin{figure}
 \includegraphics[width=\columnwidth]{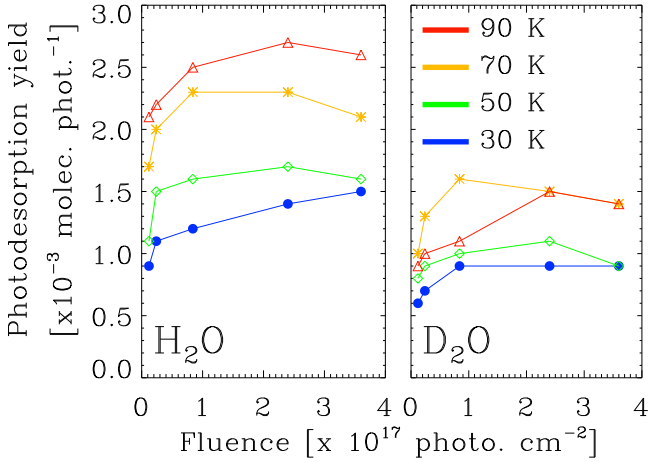}
  \caption{Photodesorption yield at different irradiation periods for temperatures of 30 K and above for water (left) and deuterated water (right).}
 \label{Fluence2}
\end{figure}

We observed the typical step-like features that correspond to the various irradiation intervals, indicating a photodesorption event from the ice mantle at every irradiation 
temperature. Notice that for temperatures of 30 K and above the photodesorption of water becomes UV-fluence dependent, see Fig. \ref{Fluence2} to appreciate better the behavior.
This could be a consequence of O$_2$ thermal desorption enhancing the photodesorption yield of H$_2$O. 

Zhu et al. (1993) summoned the MGR model to explain the difference in photodesorption between NH$_3$ and ND$_3$. They found that the photodesorption of NH$_3$ is four times 
stronger than the one for ND$_3$. It was concluded that a mass effect in the N-H coordinate was the responsible for the isotopic effect in NH$_3$ photodesorption. This model 
predicts an isotope effect stating that a lighter particle is accelerated in a shorter time on the excited potential energy surface. Therefore, this particle has a higher 
probability of escaping from the substrate. This effect has been seen in surface photon-driven processes. Adapting Eq. 2 in their publication for our experiments we obtain:
\begin{equation}
\frac{\sigma_d(H_{2}O)}{\sigma_d(D_{2}O)} = \left(\frac{1}{P_d(H_{2}O)}\right)^{\sqrt{\frac{m_{D}}{m_{H}}}-1} 
\end{equation}
where $\sigma_d(H_{2}O)$ and $\sigma_d(D_{2}O)$ are the photodesorption cross sections, in our experiment at 8 K $\frac{\sigma_d(H_{2}O)}{\sigma_d(D_{2}O)}$ 
is about 2.21 (see Table \ref{tablePhoto}), $\frac{m_{D}}{m_{H}} = 2$ for the masses of D and H atoms is an approximation of the reduced masses in the O-D and O-H stretching 
coordinates. This leads to a photodesorption probability of $P_d(H_{2}O)$ = 0.146,  i.e. nearly 15 \% of the photon-excited molecules is desorbing. The MGR model could explain the 
difference we observe between H$_2$O and D$_2$O signal intensities in the QMS after compensation from background H$_2$O contamination. H$_2$O photodesorption is two times stronger 
than the one for D$_2$O. 

\begin{table*}
 \centering
 \caption{Photodesorption yield of water, deuterated water, and its photoproducts at different irradiation temperatures. Data from different groups have been added for comparison.}
 \label{tablePhoto}
 \tiny
 \begin{tabular}{ccccc|cccccc}
  \hline
  Irrad. Temp.&\multicolumn{4}{c}{This work}&Westley et al. &Andersson &\"Oberg et al.&Arasa et al.&Arasa et al.\\
 &&&&&(1995ab)&\& van Dishoeck (2008)&(2009)&(2010)&(2011)\\
  Kelvin&H$_2$&OH&H$_2$O&O$_2$&\multicolumn{5}{c}{H$_2$O}\\
  &\multicolumn{9}{c}{$\times10^{-3}$ [$\frac{mol.}{phot.}$]}\\
  \hline
    8 &-$^*$&0.7 $\pm$ 0.3&1.3 $\pm$ 0.2&0.6 $\pm$ 0.1&&&&&\\
   10 &&&&&&0.14-0.37&&0.54&\\
   20 &&&&&&&&0.57&\\
   30 &-$^*$&0.5 $\pm$ 0.2&1.4 $\pm$ 0.4&1.1 $\pm$ 0.3&&&&0.71&\\
   35 &&&&&3.46&&&&\\
   40 &&&&&&&&&\\
   50 &-$^*$&0.6 $\pm$ 0.3&1.6 $\pm$ 0.3&1.4 $\pm$ 0.2&3.63&&&&\\
   60 &&&&&&&&&\\
   70 &-$^*$&1.0 $\pm$ 0.4&2.4 $\pm$ 0.6&2.4 $\pm$ 0.5&4.48&&&&\\
   85 &&&&&5.61&&&&\\
   90 &-$^*$&1.1 $\pm$ 0.5&2.5 $\pm$ 0.7&3.6 $\pm$ 0.9&&&&0.71&\\
  100 &&&&&7.63&&&&\\
   \hline
  \hline
  Irrad. Temp.&D$_2$&OD&D$_2$O&O$_2$&\multicolumn{5}{c}{D$_2$O}\\
  Kelvin&\multicolumn{9}{c}{$\times10^{-3}$ [$\frac{mol.}{phot.}$]}\\
  \hline
    8 & 2.7 $\pm$ 0.7&-$^*$&0.7 $\pm$ 0.1&0.5  $\pm$ 0.2&&&&&\\
   10 &&&&&&&&&1.27\\
   20 &&&&&&&2.07&&1.46\\
   30 &11.4 $\pm$ 2.2&-$^*$&0.9 $\pm$ 0.2&1.2  $\pm$ 0.2&&&2.33&&\\
   35 &&&&&&&&&\\
   40 &&&&&&&3.22&&\\
   50 &14.3 $\pm$ 2.3&-$^*$&1.0 $\pm$ 0.2&1.3  $\pm$ 0.3&&&&&\\
   60 &&&&&&&3.50&&1.57\\
   70 &18.5 $\pm$ 3.4&-$^*$&1.6 $\pm$ 0.4&2.5  $\pm$ 0.5&&&&&\\
   85 &&&&&&&&&\\
   90 &15.7 $\pm$ 3.1&-$^*$&1.5 $\pm$ 0.3&3.3  $\pm$ 0.8&&&&&1.83\\
  100 &&&&&&&4.27&&\\
  \hline
 \end{tabular}
  \\
  The Photodesorption yield measured is the average number of molecules desorbed per incident photon for the total irradiation time at selected temperature. 
 (-*) These values suffer from background H$_2$ or H$_2$O contamination in the chamber. The latter has the same m/z value of 18 as OD. Therefore, we only provide the values for 
 D$_2$ and OH, respectively.
\end{table*}

\subsection{\texorpdfstring{H$_2$/D$_2$, OH/OD, and O$_2$}{LG} photodesorption}

\begin{figure}
 \includegraphics[width=\columnwidth]{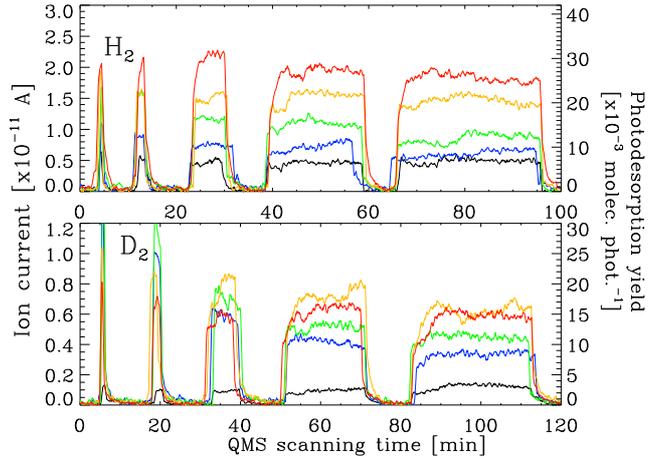}
 \caption{Photodesorption of H$_2$ (top) and D$_2$ (bottom) during irradiation of water and deuterated water ice at the 
 different irradiation doses and temperatures. The difference in QMS signal intensity is at least partially due to H$_2$ background contamination. 
 This plot follows the color code in Fig. \ref{H2O}}
 \label{H2}
\end{figure}

Hydrogen can diffuse through the ice sample at any temperature in our experiments. It is also the main contaminant inside the chamber as it is common in UHV chambers with stainless 
steel walls, contributing to the large difference between H$_2$ (m/z = 2) and D$_2$ (m/z = 4) QMS signals (H$_2$ signal almost doubles D$_2$) in Fig. \ref{H2}. This could be also 
an isotope effect as in the case of H$_2$O and D$_2$O. It can be observed that the intensity of the QMS signal is UV-fluence independent at 8 K suggesting that H$_2$/D$_2$ are formed efficiently 
and desorb very readily at a constant photon-induced desorption rate. H$_2$ can form by diffusion and reaction of H produced by photodissociation of H$_2$O giving H + OH, or 
directly from H$_2$O photodissociation into H$_2$ + O (e.g. Okabe 1978). We also observed the increase in the QMS signal with temperature, similar to water desorption data. In 
the case of H$_2$/D$_2$ a larger diffusion at higher temperatures can enhance the desorption up to almost six times higher. Since H$_2$ suffers from background contamination, only 
the D$_2$ photodesorption yield is provided in Table \ref{tablePhoto}. D$_2$ photodesorption at higher temperatures than 8 K presents a particular behavior, see Fig. \ref{H2}. D$_2$
signal presents a maximum photodesorption at the first irradiation period of 1 min. while it slowly decrease for each consequent irradiation. We are not sure about what is causing
this behavior. Molecular hydrogen production by UV-irradiation using pure methanol or pure water ice is remarkably different. 
The H$_2$ photodesorption in water is three orders of magnitude lower in this work compared to Cruz-Diaz et al. (2016) for methanol. 

\begin{figure}
 \includegraphics[width=\columnwidth]{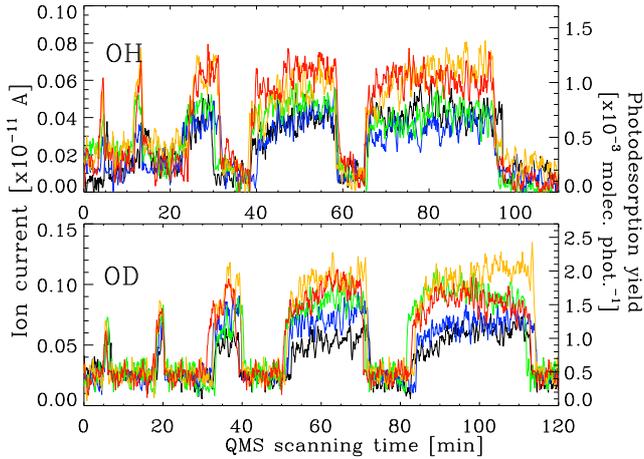}
 \caption{Photodesorption of OH (top) and OD (bottom) during irradiation of water and deuterated water ice at the different irradiation doses and temperatures. The data have been 
 smoothed to help in the visualization of the photodesorption events. This plot follows the color code in Fig. \ref{H2O}}
 \label{OH}
\end{figure}

During H$_2$O/D$_2$O ice irradiation, OH (m/z = 17) and OD (m/z = 18) photodesorption is observed besides their low intensity in the QMS signal (data was smoothed for a better 
appreciation of the step-like features), see Fig. \ref{OH}. Only the OH photodesorption yield is given in Table \ref{tablePhoto} since OD signal could be contaminated by 
background water since they share the m/z = 18. We took into account the portion of H$_2$  and OH signals coming from the dissociation of H$_2$O inside the QMS apparatus using 
the ratio in Fig. \ref{masas}. Because the fragmentation ratio of H$_2$O on the filament of the QMS may vary between experiments (this is an empirical result, it is a behavior we 
have noticed during the time the QMS have been active. We cannot explain it but we can monitor the issue and know when it happens therefore we can corrected), the values provided 
for D$_2$ and OH have a significant error. OH photodesorption is UV-fluence independent during irradiation at 8 K.

\begin{figure}
 \includegraphics[width=\columnwidth]{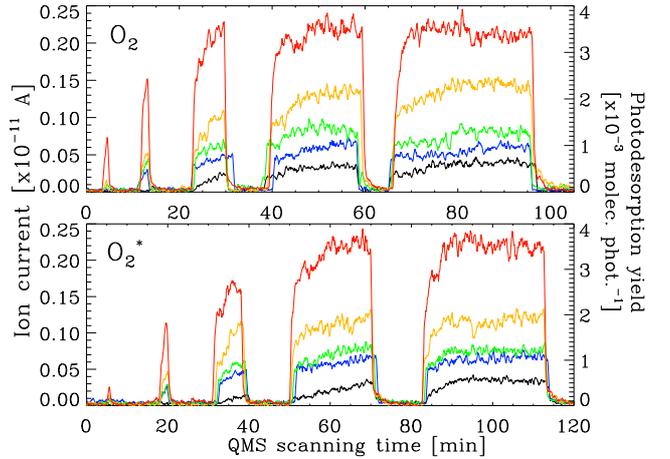}
 \caption{Photodesorption of O$_2$ at different irradiation doses and temperatures. It presents the step like function progressively increasing with irradiation time, 
 denoting a bulk effect and a UV-fluence dependence. (*) Oxygen QMS signal during D$_2$O irradiation. This plot follows the color code in Fig. \ref{H2O}}
 \label{O2}
\end{figure}

Molecular oxygen, O$_2$, has a strong QMS signal during its photodesorption, see Fig. \ref{O2}. It reaches the same intensity in both experiments (H$_2$O and D$_2$O irradiation), 
which increases progressively with irradiation time up to a maximum, denoting a bulk and a fluence dependence effect. This explains the fluence dependence of molecular oxygen 
photodesorption plotted in Fig \ref{correction} bottom (see Cruz-Diaz et al. 2016 for a detailed description of this effect). This trend in the 
photodesorption of O$_2$ is expected because O$_2$ is not formed by a primary photodissociation process of H$_2$O. At high ice temperatures, diffusion of O atoms is enhanced and 
O$_2$ photodesorption increases. It is worth to notice the kinetic observed with the light on/off intervals follow that of a continuous irradiation over the explored fluence 
range. We observed that the end of an irradiation coincides with the start of the other in the QMS signal intensity, see Fig. \ref{O2} for a clear example.

Atomic oxygen, H$_2$O$_2$, and O$_3$ were not found to desorb upon UV-irradiation in our experiments. We found that the main contribution to m/z =16 of O$^+$ was due to 
H$_2$O/D$_2$O and O$_2$ fractionation in the QMS filament. Therefore, based on our data it was not possible to infer the photodesorption of oxygen atoms. We found no evidence for 
H$_2$O$_2$ or O$_3$ formation. The former was not seen in IR or QMS data (m/z = 34) which indicates its abundance is too low. Ozone (m/z = 48) IR features were not observed neither 
its QMS signal. This could be because, if produced, it would be dissociated into O$_2$ + O.

\section{Discussion}\label{sec:discussion}

Many attempts to measure the photodesorption of water have been published over the past years. It is such a hot topic mainly because of the implications of water photodesorption in 
astronomical research. This work is justified by the current disagreement between experimental values from different teams. Our method, after calibration of our QMS, can give the 
total amount of photodesorbing molecules directly measured in the gas-phase, and not indirectly in the solid phase. Here we compare our results to those of selected works previously 
published, all of them indirectly measuring the photodesorption yield. 

OH has been detected toward Sgr B2 with an abundance of 2 to 5 $\times$ 10$^{-6}$, Goicoechea \& Cernicharo (2002). They attributed this OH abundance to a photon-dominated 
region. Our experiments give proof that photons can be responsible for the photodesorption of this molecule, in line with other experimental works like \"Oberg et al. (2009).

Andersson \& van Dishoeck (2008) used classical Molecular Dynamics methods with analytical potentials to calculate the photodissociation and photodeorption of water ice. They 
showed that water is more likely to be dissociated than to photodesorb, hence the discrepancy in the water photodesorption measurements found in the literature. They distinguished 
three kinds of mechanism for water photodesorption: kick out of a water molecule by a hot H atom, desorption by the excess of energy during recombination of H and OH, and the 
desorption of H and OH separately. They reported no evidence of photodesorption below the fifth monolayer. Assuming that water molecules detected in the QMS are photodesorbing, we 
calculated the photodesorption yield per absorbed photon for the first five monolayers using equation 1 and the average VUV absorption cross-section of solid H$_2$O and D$_2$O, 
see Table \ref{Abs}. We cannot distinguish by monolayer nor by mechanism hence we show the total water photodesorption.

Andersson \& van Dishoeck (2008) affirmed that the more mobility of the photoproducts the more probable they will react with other species. Also, that even they did not include 
O atoms in their calculations, photoproducts like O$_2$ could be photodesorbing from the ice. These photoproducts, and the fact that mobility is increased by temperature, 
could explain the increase of water photodesorption with temperature and its fluence dependence.

Theoretical calculations of Andersson \& van Dishoeck (2008) predicted a total photodesorption yield of 5.1 $\times$ 10$^{-4}$ absorbed photons$^{-1}$, which is two orders of magnitude 
lower than our result at 8 K, 7 $\times$ 10$^{-2}$ molecules per absorbed photon. We are not sure why this difference but QMS data show a higher photodesorption of water related to OH 
where Andersson \& van Dishoeck (2008) results show the opposite. Speculating, this could be an effect of the UV field spectrum (UV wavelength coverage of the lamp) and/or the fact 
that photoproducts different than H and OH are photodesorbing and could help in the photodesorption of water by the kick out process and/or lowering the binding energy of the water 
molecule about to photodesorb. However this have to be tested by models.

\begin{table}
  \centering
  \caption{Total photodesorption yield per absorbed photon at different temperatures.} 
  \begin{tabular}{ccc}
  \hline
  \hline
Temp.& H$_2$O& D$_2$O\\
  (K)&  \multicolumn{2}{c}{(molec. photon$^{-1}$)}\\
    \hline
  8 &0.07&0.05\\
 30 &0.08&0.06\\
 50 &0.09&0.07\\
 70 &0.14&0.12\\
 90 &0.15&0.11\\
 \hline
  \end{tabular}
 \label{Abs}
\end{table}

Molecular dynamics simulations by Arasa et al. (2010, 2011), studied the UV photodissociation of amorphous water and deuterated water ice at 10, 20, 30, and 90 K analyzing 
the effect of ice temperature. They found an increasing in water photodesorption of up to 31 \% for H$_2$O and 44 \% for D$_2$O at 90 K. In our case, we found the similar behavior 
but with a higher percentages, 92 and 114 \%, respectively, at 90 K. Comparing their photodesorption yields with Table \ref{Abs} a discrepancy of more than two orders of magnitude is 
observed. As with Andersson \& van Dishoeck (2008) photodesorption yields, this could be an effect of the photodesorbing photoproducts. Nevertheless, Arasa et al. (2010, 2011) had 
short time scales in their simulations, where thermally activated processes like diffusion and thermal desorption were not probed, arguing that if longer time scales are considered, 
a stronger dependence on ice temperature would be expected.

Watanabe et al. (2000) focused on the production of D$_2$ by the UV-irradiation of D$_2$O ice mantles at 12 K and temperatures above it. They found that D$_2$ desorption increase 
drastically for temperatures above 20 K, this corresponds to the evaporation temperature of D from the ice (Laufer et al. 1987). This is in agreement with our results, we also detected 
this increasing between 8 to 30 K, see Fig. \ref{H2}. Watanabe et al. (2000), after a dose of 10$^{18}$ photons cm$^{-2}$, measured a production of D$_2$ of about 1 - 2 \% of the 
initial number of D$_2$O molecules at 12 K. In our case, taking the ice thickness for D$_2$O at 8 K in Table \ref{exp} (245 ML) and the total number of photodesorbed D$_2$ molecules 
(1.94 ML) for the same experiment, we have a total of $\sim$ 1\%. In agreement with Watanabe et al. (2000).

The UV filed experienced by the ice mantles in a dark cloud is induced by cosmic rays and is about 10$^4$ photons cm$^{-2}$ s$^{-1}$, see Gredel et al (1989) and Shen et al. (2004). 
Using this UV flux in equation 1, accounting for the top five monolayers, the absorbed photons can be calculated. Taking the photodesorption yield for H$_2$O in Table \ref{Abs} at 
8 K, we calculated a total water desorption rate of $\sim$ 12 molecules cm$^{-2}$ s$^{-1}$. For a dark cloud with a mean life time of 1 Myr, the total photodesorbed water would be 
3.8 $\times$ 10$^{14}$ molecules cm$^{-2}$ $\sim$ 0.4 ML. This is a low number compared to gas-phase water abundance toward massive protostars $\sim$ 10$^{18}$ molecules cm$^{-2}$ 
(Boonman et al. 2000 and Boonman \& van Dishoeck 2003) but in cold clouds, Zmuidzinas et al. (1995) and Tauber et al. (1996) have found that the gas-phase H$_2$O abundance is 
low, $\sim$ 10$^{-8}$ to 10$^{-7}$ related to H$_2$.

Yeghikyan (2017), using the Cloudy code (Ferland et al. 2013), calculated the effect of the irradiation on the abundance of water (ice and gas) in planetary nebulae (PN) for 
1000 yr. They concluded that this abundance depends on the ionization rate of hydrogen by energetic particles like cosmic rays (CR). They calculated the VUV flux inside the PN, on 
average 1.6 $\times$ 10$^{-3}$ erg cm$^{-2}$ s$^{-1}$ with an energy per photon close to 10 eV. This gives an average photon flux of 1 $\times$ 10$^8$ photons cm$^{-2}$ s$^{-1}$. 
Taking this flux and making the same calculations as before, we found an average water photodesorbed column density of 3.7 $\times$ 10$^{15}$ molecules cm$^{-2}$ in 1000 yr. This 
result is in line with Yeghikyan (2017) for a CR rate of 10$^{-13}$ - 10$^{-14}$ s$^{-1}$ in their models.

Abdulgalil et al. (2017) bombarded water ice with low-energy electrons. They found similar results by detecting the photodesorption of H$_2$O/D$_2$O as well as the photoproducts 
H$_2$/D$_2$ and O$_2$. As in our case, their major photoproduct by far was hydrogen followed by oxygen.

Westley et al. (1995a,b) deposited a 0.5 microns ice, a much thicker ice compared to ours, this could justify the difference in photoproduct formation. They used a hydrogen lamp 
with a strong Ly-$\alpha$ emission and negligible molecular hydrogen emission around 7.8 eV (besides Ly-$\alpha$, our lamp presents this bands). They gave a rage for the photon flux, 
from 0.5 to 5 $\times$ 10$^{14}$ photons/cm$^2$ s (ours has been well constrain to 2 $\times$ 10$^{14}$ photons/cm$^2$ s). In addition, their irradiation dose is much higher than ours 
($\sim$ 6 $\times$ 10$^{18}$ compared with 3.6 $\times$ 10$^{17}$, about 16 times higher). This could be because they needed a higher dose to remove enough water to be able to measure 
the thickness more accurately. Nevertheless, we obtained similar results once they are calibrated and properly compared. In their UV irradiation experiments, Westley et al. (1995a,b) 
observed an incubation stage with incubation dose of $\sim$ 10 $\times$ 10$^{18}$ photons/cm$^2$. This suggests a dependence of the photodesorption yield with the UV-fluence. Their 
irradiation experiments are at temperatures above 30 K. Above this temperature H$_2$ and O$_2$ can move with a higher freedom than at 8 K as shown in Figs. \ref{H2} and \ref{O2}, 
increasing in the QMS signal with each irradiation period. For these temperatures water photodesorption gets UV-fluence dependent. Fig. \ref{Fluence2} resembles Fig. 2 in Westley 
et al. (1995b), meaning that their incubation dose corresponds to the higher diffusion of H$_2$ and O$_2$ which helps water to photodesorb. Their dependency with UV-fluence is in 
agreement with our results. 

\begin{figure}
 \includegraphics[width=\columnwidth]{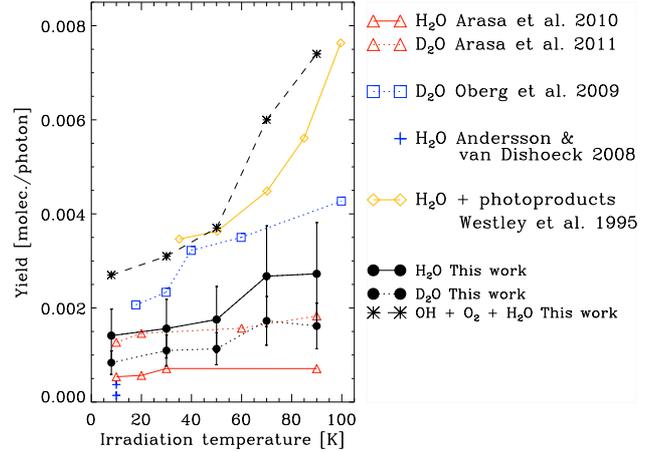}
 \caption{Photodesorption yield of water and deuterated water found by other groups compared with our work.}
 \label{Yield}
\end{figure}

Westley et al. (1995a,b) used quartz crystal microbalance to measure the photodesorption yield of H$_2$O, therefore the yield they derived included all photodesorbing species, 
namely H$_2$, OH, H$_2$O and O$_2$ . This statement is briefly mentioned by Andersson \& van Dishoeck (2008), stating that the removal of H$_2$O from the surface was driven by H 
and OH desorption in the top two monolayers. In Fig. \ref{correction}, if we add yields of H$_2$, OH and O$_2$ together (like in Fig. \ref{Yield}), the trend of total desorbing 
molecules/radicals will be fluence-dependent which agrees with Westley et al. (1995a,b) report, because quartz crystal microbalance can only monitor total loss yield, it cannot 
distinguish the proportion of desorbing molecules/radicals. In contrast, our study can provide photodesorption yield of each measurable desorbing species a from H$_2$O ice mantle.

Using the yields shown in Table \ref{tablePhoto}, we compared our photodesorption of water with Westley et al. (1995a) by adding the 
photodesorption yields of OH, H$_2$O, and O$_2$, please confront black asterisks with yellow diamonds in Fig. \ref{Yield}. The photodesorption increases with temperature with 
a rate similar to Westley et al. (1995a), although H$_2$ was not added (since it is affected by the isotopic effect explained above in the case of H$_2$O and D$_2$O) which means the 
behavior could be different. Although we get similar results as Westley our analytical tools are different. Photolysis becomes increasingly important as temperature increase. 
This suggests a gas-phase water gradient that increases toward hotter parts in a protoplanetary disk having a direct impact on the position and extension of the water snow line and 
ultimately affecting the planetary formation and the habitable zone, see Cieza et al. (2016). In addition to photodesorption, cosmic ray sputtering is expected to contribute 
significantly to the release of molecules from ice mantles in cold inter- and circumstellar regions (Dartois et al. 2013).

\"Oberg et al. (2009) reported the photodesorption yield for D$_2$O using the RAIR spectra of its stretching band at different temperatures. This technique has the disadvantage 
that it has no direct relation to column density and it is affected by the OH production. In one hand, they calculated a photodesorption yield $\sim$ 3 times higher compared with 
our D$_2$O photodesorption yield, with a total 60 \% error in their calculations. In the other hand, their dependence with temperature is similar to ours, an almost linear increment. 
The discrepancy between the two yields is probably due to the fact that we measured the D$_2$O photodesorption yield directly from the QMS, while in \"Oberg et al. (2009) IR 
spectroscopy is used, thus leading to an indirect calculation of the D$_2$O photodesorption yield, since both photodissociation and photodesorption are taken into account in that 
case. Their H$_2$O to OH ratio is 1 at low temperatures and 2 at high temperatures. We obtained ratios of 2 and 2.5 at 8 and 90 K. The primary difference is the techniques used by 
the two works. We consider RAIR spectroscopy a powerful tool but it fails to quantify the column density of the deposited sample, a key parameter to determine a photodesorption yield.

De Simone et al. (2013) addresses the water photodesorption problem by irradiating water ice at 108 K using a monochromatic light with a wavelength centered at 157 nm. They used 
time-of-flight mass spectrometer and resonance-enhanced multiphoton ionization to measure the photodesorption cross section of water at 108 K, on average 6.9 $\pm$ 1.8 $\times$ 
10$^{-20}$ cm$^2$ for $>$ 10 Langmuir water exposure. They did not provide a photodesorption yield because their relation between water signal and thickness was unknown. 
Nevertheless, they estimated a yield of 1.8 $\times$ 10$^{-4}$ molecules per photon. Being this a yield value obtained at high temperature we can only compare it with our photodesorption 
yield at 90 K, 2.5 $\times$ 10$^{-3}$ molecules per photon for H$_2$O and 1.5 $\times$ 10$^{-3}$ molecules per photon for D$_2$O. DeSimone et al (2013) estimation is then one order of 
magnitude lower than ours, \"Oberg et al. (2009), and Arasa et al. (2011). This is a similar problem as above, knowing the column density is crucial for measuring the photodesorption 
yield. 

\section{Conclusion}\label{sec:conclusion}

Photodesorption yields have been given using a novel approach. Based on the experiments presented in this work, we estimated the photodesorption rates at different temperatures 
for H$_2$O/D$_2$O, along with H$_2$/D$_2$, OH/OD and O$_2$ photoproducts, see Table \ref{tablePhoto}. In the case of O$_2$, the photodesorption increases with irradiation time, a 
process highly influenced by the building up and diffusion of O$_2$ molecules in the bulk of the ice sample which eventually reach the surface and desorb. This denotes a fluence 
dependence in the photodesorption of O$_2$ molecules. This diffusion affects directly other molecules like H$_2$O, increasing the photodesorption of it. 
The photodesorption effect can explain the presence of gas-phase water in cold regions where thermal processes are inhibited. Data reported 
here can be used as inputs for numerical simulations of those regions. We compared our results with existing data and gave possible explanations for the discrepancies. Comparison 
between our work and previous theoretical simulations is provided for an informational purpose only, since this is beyond the scope of this work.

\section*{Acknowledgements}

This research was financed by the Spanish MINECO under project AYA2011-29375, AYA2014-60585-P. E. Moreno was also financed by CONSOLIDER grant CSD2009-00038. This work was partially 
supported by MOST103-2112-M-008-025- MY3 (Y-JC), Taiwan.


\appendix

\section{Data reduction}

In this section, we will explain in detail the steps taken to reduce the raw data from the QMS presented here. We will not explain however, the calibration of the QMS because as 
mentioned in the Methods section, it was already explained in a previous paper, see Mart\'\i{}n-Dom\'enech et al. (2015).

\begin{figure}
 \includegraphics[width=\columnwidth]{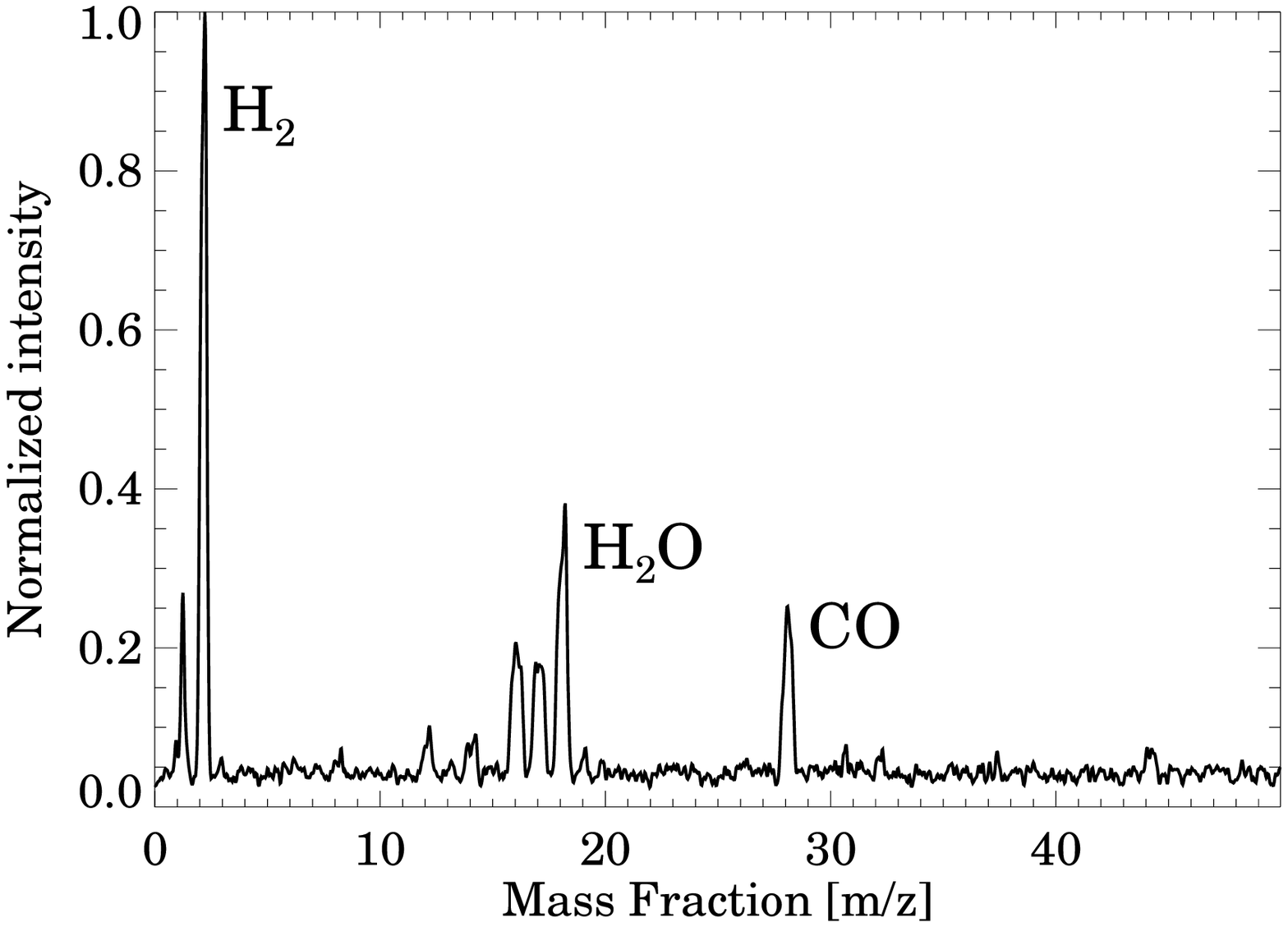}
 \caption{Contamination inside ISAC. Spectrum taken at 8 K and normalized by the most intense mass fragment: m/z = 2.}
 \label{chamber}
\end{figure}

\begin{figure}
 \includegraphics[width=\columnwidth]{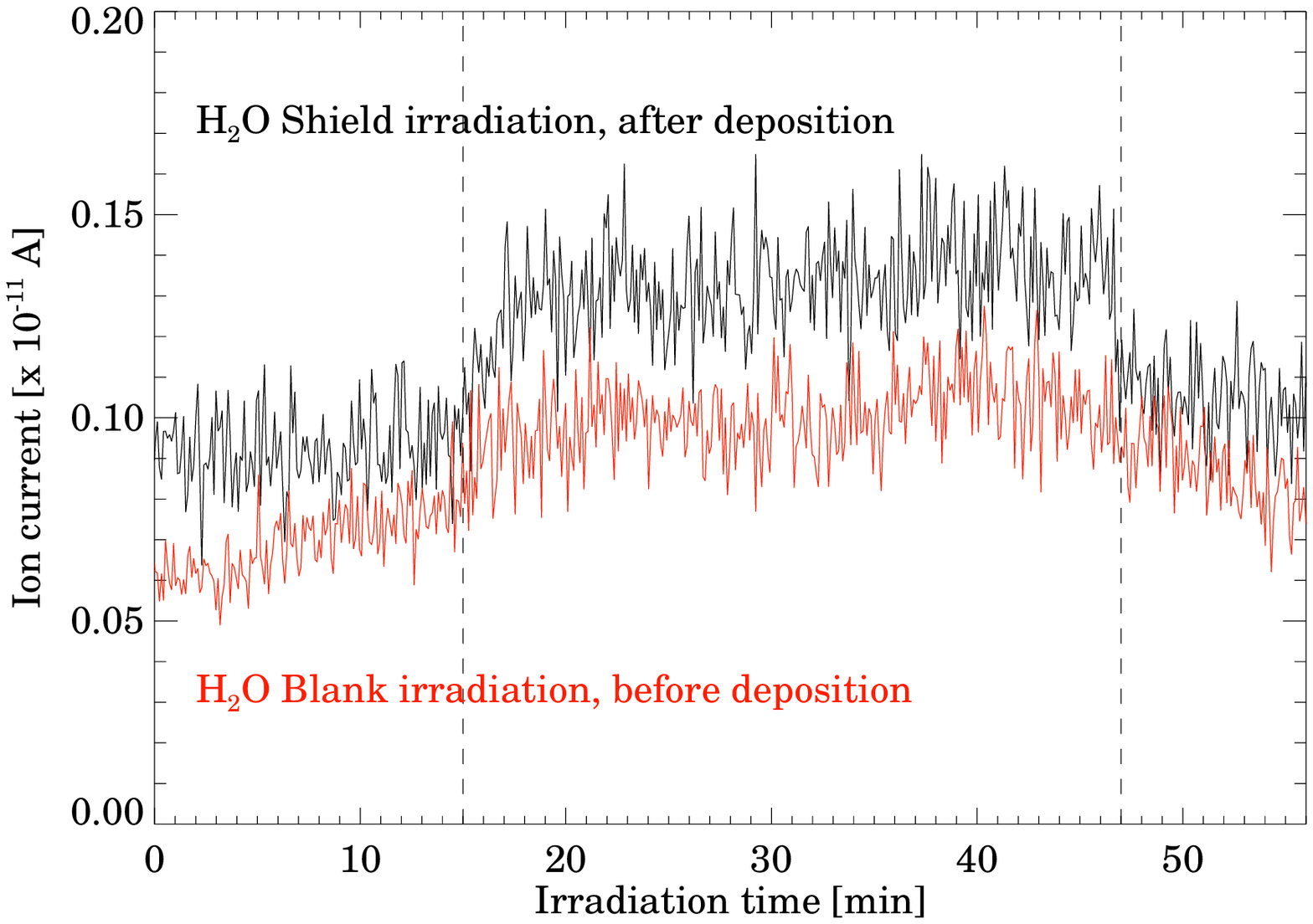}
 \caption{Raw data showing the irradiation test to detect the background water contamination in the experiments. Dashed line shows the beginning and the end of the irradiation 
 period for the two test irradiation.}
 \label{test1}
\end{figure}

The main contaminant inside the UHV chamber is hydrogen, followed by water and carbon monoxide, see Fig. \ref{chamber}. Aiming to correct the data from background water contamination, 
we carried out two irradiations of half an hour before starting the UV-photolysis of the ice sample. First, at 8 K, we irradiated the KBr window with no deposition. Then after 
deposition, we turned the cold finger 90$^{\circ}$ to irradiate its shield. During deposition most of the water vapor will condensate on top of the window but a small amount of water 
molecules will end up sticking elsewhere. These irradiation gave us two different signals for the possible contamination coming from background water, see Fig. \ref{test1}. On average,
water contamination coming from the blank irradiation has an intensity of 3 $\times$ 10$^{-13}$ Amps in the QMS while the irradiation of the shield reaches an average intensity of 
5 $\times$ 10$^{-13}$ Amps, after baseline correction (see below). Also, we used Fig. \ref{masas} to remove the contribution of the fragmentation of H$_2$O and D$_2$O in the 
filaments of the QMS. For H$_2$/H$_2$O fragmentation ratio we found a contribution of 1 \%, 0.7 \% for D$_2$/D$_2$O, 37 \% for OH/H$_2$O, and 56 \% for OD/D$_2$O. We decided to sum 
the contributions and subtract the result from the water irradiation experiments. This method was used to correct the m/z signal coming from H$_2$/D$_2$ and OH/OD photodesorbed 
from the ice as well.

Data in Figs. 3, 5, 6, and 8 were corrected by baseline. This means, we brought to zero the QMS signal for clarity, Fig. \ref{test1} is an example of raw data. Nevertheless, 
quantifications were performed using the raw data after water subtraction, as explained above.

Error estimations were calculated using measurements of the average QMS noise, extracted from the raw data, and the uncertainty in the column density calculation, $N$, using 
equation 2. Using the variance formula for the error propagation, we have:
\begin{equation}
\delta(Y) =  \sqrt{\frac{\delta_i^2 + \delta_j^2 + \delta_k^2 + ... + \delta_n^2}{n-1}} .
\end{equation}
being $\delta(Y)$ the error in the photodesorption yield and $\delta_n$ the number of possible errors to be considered as the calculation of the column density, UV-flux, etc.
$N$, the column density, itself is defined by the error propagation of equation 2 and by the QMS signal noise, which sometimes can be as intense as the measured signal. In the case 
of the UV-flux it depends of the method used to derive it. Mu\~{n}oz Caro et al. (2010) has a full appendix on how they derived this value. Also, the lamp fluctuations noticed in 
the QMS data. The latter was caused by variations on the H$_2$ pressure inside the UV-lamp. Co-adding data would work for instance in the acquisition of IR data, but it is not a 
valid option in this case since the QMS signal and the lamp fluctuations vary from one day to another, it adds uncertainty. H$_2$O photodesorption measurements in the gas phase are 
difficult to obtain hence we improved as much as we could to obtain optimum results.


\bsp	
\label{lastpage}

\begin{thebibliography}{99}
\bibitem[\protect\citeauthoryear{Abdulgalil}{2017}]{Abdulgalil}
Abdulgalil, A. G. M., Rosu-Finsen, A., Marchione, D., et al., 2017, ACS Earth Space Chem., 1, 209
\bibitem[\protect\citeauthoryear{Akinori}{2012}]{Akinori} 
Akinori O., Taishi N., \& Shigeru I., 2012, ApJ, 738, 2 
\bibitem[\protect\citeauthoryear{Allamandola}{1988}]{Allamandola} 
Allamandola L. J., Sandford S. A., \& Valero G. J., 1988, Icarus, 76, 225
\bibitem[\protect\citeauthoryear{Andersson}{2008}]{Andersson} 
Andersson S., \& van Dishoeck E. F., 2008, A\&A, 491, 3, 907
\bibitem[\protect\citeauthoryear{Arasa1}{2010}]{Arasa1} 
Arasa C., Andersson S., Cuppen H. M., van Dishoeck E. F. \& Kroes G. -J., 2010, Journal of Chemical Physics, 132, 18, 184510
\bibitem[\protect\citeauthoryear{Arasa2}{2011}]{Arasa} 
Arasa C., Andersson S., Cuppen H. M., van Dishoeck E. F. \& Kroes G. -J., 2011, Journal of Chemical Physics, 134, 16, 164503
\bibitem[\protect\citeauthoryear{Banzatti}{2008}]{Banzatti} 
Banzatti A., Pinilla P., Ricci L., et al., 2015, ApJL, 815, L15
\bibitem[\protect\citeauthoryear{Blum}{2008}]{Blum} 
Blum J. \& Wurm G., 2008, Annu. Rev. Astron. Astrophys., 46, 21
\bibitem[\protect\citeauthoryear{Bertin}{2013}]{Bertin} 
Bertin M., Claire R., Mikhail D., et al., 2016, ApJ, 817, 2, L12  
\bibitem[\protect\citeauthoryear{Boonman}{2000}]{Boonman}
Boonman, A. M. S., van Dishoeck, E. F., Lahuis, F., Wright, C. M., \& Doty, S. D., 2000, ISO beyond the peaks: The 2nd ISO workshop on analytical spectroscopy. Eds. A. Salama, 
M.F.Kessler, K. Leech \& B. Schulz. ESA-SP 456. p.67
\bibitem[\protect\citeauthoryear{Boonman}{2003}]{Boonman2}
Boonman, A. M. S. \& van Dishoeck, E. F., 2003, A\&A, 403, 1003
\bibitem[\protect\citeauthoryear{Celiberto}{2005}]{Celiberto} 
Celiberto R., Janev R. K., Laricchiuta A., et al., 2001, Atomic Data and Nuclear Data Tables 77 161
\bibitem[\protect\citeauthoryear{Cernicharo}{2005}]{Cernicharo} 
Cernicharo J. \& Crovisier J., 2005, Space Science Reviews, 119, 1-4, 29
\bibitem[\protect\citeauthoryear{Chen}{2014}]{Chen} 
Chen Y.-J., Chuang K.-J., Mu\~noz Caro G. M., et al., 2014, ApJ, 781, 1, 15
\bibitem[\protect\citeauthoryear{Cieza}{2016}]{Cieza} 
Cieza L. A., Casassus S., Tobin J., et al., 2016, Nature, 535, 258
\bibitem[\protect\citeauthoryear{Collings}{2004}]{Collings} 
Collings M. P., Anderson M. A., Chen R., et al., 2004, Mon. Not. R. Astron. Soc., 354, 1133
\bibitem[\protect\citeauthoryear{Cruz-Diaz}{2014a}]{Cruz-Diaza} 
Cruz-Diaz G. A., Mu\~noz Caro G. M., Chen Y.-J., \& Yih T.-S., 2014, A\&A, 562, A119
\bibitem[\protect\citeauthoryear{Cruz-Diaz}{2014b}]{Cruz-Diazb} 
Cruz-Diaz G. A., Mu\~noz Caro G.M., \& Chen Y.-J., 2014, MNRAS, 430, 2370
\bibitem[\protect\citeauthoryear{Cruz-Diaz}{2016}]{Cruz-Diaz} 
Cruz-Diaz G .A., Mart\'\i{}n-Dom\'enech R., Mu\~noz Caro G. M., \& Chen Y.-J., 2016, A\&A, 592, A68
\bibitem[\protect\citeauthoryear{Dartois}{2013}]{Dartois} 
Dartois E., Ding J. J., de Barros A. L. F., et al., 2013, A\&A, 557, A97 
\bibitem[\protect\citeauthoryear{Deutsch}{2013}]{DeSimone} 
DeSimone A. J., Crowell V. D., Sherill C. D., \& Orlando T. M., 2013, J. Chem. Phys., 139, 164702
\bibitem[\protect\citeauthoryear{Deutsch}{1997}]{Deutsch} 
Deutsch H., Becker K., \& M\"ark T. D., 1997, Int. J. Mass. Spectrom. Ion Processes, 168, 503
\bibitem[\protect\citeauthoryear{Ferland}{2013}]{Ferland}
Ferland G., Porter R., van Hoof P., et al., 2013, RMXAA, 49, 137 .
\bibitem[\protect\citeauthoryear{Gibb}{2004}]{Gibb} 
Gibb E. L., Whittet D. C. B., Boogert A. C. A., \& Tielens A. G. G. M., 2004, ApJ Supplement Series, 151, 1, 35
\bibitem[\protect\citeauthoryear{Goicoechea}{2002}]{Goicoechea} 
Goicoechea J. R. \& Cernicharo J., 2002, ApJ, 576, L77
\bibitem[\protect\citeauthoryear{Gredel}{1989}]{Gredel} 
Gredel, R., Lepp, S., \& Dalgarno, A. 1989, ApJ, 347, 289
\bibitem[\protect\citeauthoryear{Greenberg}{1980}]{Greenberg} 
Greenberg J. M., Allamandola L. J., Hagen W., van de Bult C. E. P., \& Baas F., 1980, Proceedings of the Symposium, Mont Tremblant, Quebec, Canada, 355
\bibitem[\protect\citeauthoryear{Fayolle}{2011}]{Fayolle} 
Fayolle E. C., Bertin M., Romanzin C., et al., 2011, ApJ Letters, 739, L36
\bibitem[\protect\citeauthoryear{Hagen}{1981}]{Hagen} 
Hagen W., Tielens A. G. G. M., \& Greenberg J. M., 1981, Chemical Physics, 56, 367
\bibitem[\protect\citeauthoryear{Honda}{2016}]{Honda} 
Honda M., Kudo T., Takatsuki S., et al., 2016, ApJ, 821, 1, 2
\bibitem[\protect\citeauthoryear{Howard}{2013}]{Howard} 
Howard C. D., Sandell G., Vacca W. D., et al., 2013, ApJ, 776, 21
\bibitem[\protect\citeauthoryear{Krishnakumar}{1992}]{Krishnakumar} 
Krishnakumar E. \& Srivastava S. K., 1992, Int. J. Mass. Spectrom. Ion Processes, 113, 1 
\bibitem[\protect\citeauthoryear{Laufer}{1987}]{Laufer} 
Laufer, D., Kochavi, E., \& Bar-Nun, A., 1987, Phys. Rev. B, 36, 9219
\bibitem[\protect\citeauthoryear{Lindberg}{2014}]{Lindberg} 
Lindberg J. E., Jorgensen J. K., Green J. D., et al., 2014, A\&A, 565, A29
\bibitem[\protect\citeauthoryear{Mart\'\i{}n-Dom\'enech}{2014}]{Rafa1} 
Mart\'\i{}n-Dom\'enech R., Mu\~noz Caro G. M., Bueno J., \& Goesmann F., 2014, A\&A, 564, A8
\bibitem[\protect\citeauthoryear{Mart\'\i{}n-Dom\'enech}{2015}]{Rafa2} 
Mart\'\i{}n-Dom\'enech R., Manzano-Santamar\'\i{}a J., Mu\~noz Caro G. M., et al., 2015, A\&A, 584, A14
\bibitem[\protect\citeauthoryear{Meeus}{2010}]{Meeus} 
Meeus G., Pinte C., Woitke P., et al., 2010, A\&A, 518, L124
\bibitem[\protect\citeauthoryear{Mu\~noz Caro}{2003}]{Caro2} 
Mu\~noz Caro G. M., \& Schutte W. A., 2003, A\&A, 412, 121
\bibitem[\protect\citeauthoryear{Mu\~noz Caro}{2010}]{Caro1} 
Mu\~noz Caro G. M., Jim\'enez-Escobar, A., Mart\'\i{}n-Gago J. \'A., et al., 2010, A\&A, 522, A108
\bibitem[\protect\citeauthoryear{\"Oberg}{2010}]{Oberg} 
\"Oberg K. I., van Dishoeck E. F., Linnartz H., \& Andersson S., 2010, ApJ, 718, 832 
\bibitem[\protect\citeauthoryear{Okabe}{1978}]{Okabe} 
Okabe H., 1978, Photochemistry of small molecules, ed. John Wiley \& Sons, New York
\bibitem[\protect\citeauthoryear{Pilbratt}{2010}]{Pilbratt} 
Pilbratt G. L., Riedinger J. R., Passvogel T., et al., 2010, A\&A, 518, L1 
\bibitem[\protect\citeauthoryear{Rapp}{1965}]{Rapp} 
Rapp D. \& Englander-Golden P., 1965., J. Chem. Phys. 43 1464
\bibitem[\protect\citeauthoryear{Riviere}{2015}]{Riviere} 
Riviere-Marichalar P., Bayo A., Kamp I., et al., 2015, A\&A, 575, A19
\bibitem[\protect\citeauthoryear{Riviere}{2016}]{Riviere2} 
Riviere-Marichalar P., Mer\'\i{}n B., Kamp I., Eiroa C., \& Montesinos B., 2016, A\&A, 94, A59
\bibitem[\protect\citeauthoryear{Shen}{2004}]{Shen}
Shen, C., Greenberg, J. M., Schutte, W. A., \& van Dishoeck, E. F., 2004, A\&A, 415, 203
\bibitem[\protect\citeauthoryear{Straub}{1998}]{Straub} 
Straub H. C., Lindsay B. G., Smith K. A., \& Stebbings R. F., 1998, J. Chem. Phys., 108, 109
\bibitem[\protect\citeauthoryear{Tauber}{1996}]{Tauber} 
Tauber J., Olofsson G., Pilbratt G., Nordh L., \& Frisk U., 1996, A\&A, 308, 913
\bibitem[\protect\citeauthoryear{Thi}{2013}]{Thi} 
Thi W. F., M\'enard F., Meeus G., et al., 2013, A\&A, 557, A111
\bibitem[\protect\citeauthoryear{Tilling}{2012}]{Tilling} 
Tilling I., Woitke P., Meeus G., et al., 2012, A\&A, 538, A20
\bibitem[\protect\citeauthoryear{Watanabe}{2000}]{Watanabe}
Watanabe, N., Toshikazu, H., \& Kouchi, A., 2000, ApJ, 541, 772
\bibitem[\protect\citeauthoryear{Westley}{1995a}]{Westley} 
Westley M. S., Baragiola R. A., Johnson R. E., \& Baratta G. A., 1995a, Nature, 373, 6513, 405
\bibitem[\protect\citeauthoryear{Westley}{1995b}]{Westley2} 
Westley M. S., Baragiola R. A., Johnson R. E., \& Baratta G. A., 1995b, Planet. Space Sci., 43, 10/11, 1311
\bibitem[\protect\citeauthoryear{Yeghikyan}{2017}]{Yeghikyan}
Yeghikyan, A, 2017, Molecular Astrophysics, 8, 40
\bibitem[\protect\citeauthoryear{Yoon}{2010}]{Yoon} 
Yoon J.-S., Kim Y.-W., Kwon D.-C., et al., 2010, Reports on Progress in Physics, 73, 116401
\bibitem[\protect\citeauthoryear{Zhu}{1993}]{Zhu} 
Zhu X.-Y., Wolf M., Huett T., \& White J. M., 1993, Desorption Induced by Electronic Transitions (DIET V), eds. A. R. Burns, E. B. Stechel, D. R. Jennison (Springer-Verlag), 63
\bibitem[\protect\citeauthoryear{Zmuidzinas}{1995}]{Zmuidzinas}
Zmuidzinas J., Blake G. A., Carlstrom J., Keene J., Miller D., et al., 1995, Proc. Airborne Astron. Symp., ed. MR Haas, JA Davidson, EF Erickson, San Francisco: Astron. Soc. Pac., 33-40. 
\bibitem[\protect\citeauthoryear{Zubko}{2004}]{Zubko}
Zubko, V., Dwek, E., \& Arendt, R. G., 2004, ApJ Supplement Series, 152, 211 
\end{thebibliography}
\end{document}